\documentstyle[12pt]{article}
\renewcommand{\theequation}{\arabic{section}.\arabic{equation}}

\begin{document}

\begin{flushright}
UT-Komaba 98-20 \\
ICRR-Report-432-98-28
\end{flushright}

\begin{center} 
{\Large{\bf  Nonlocally-Correlated Disorder and Delocalization in One Dimension: }}\\
\vskip 0.15cm
{\Large{\bf Density of States }}
\vskip 1.5cm

{\Large  Ikuo Ichinose$^{\dagger}$\footnote{e-mail 
 address: ikuo@hep1.c.u-tokyo.ac.jp}and Masaomi Kimura$^{\ast}$\footnote{e-mail
 address: masaomi@ctpc1.icrr.u-tokyo.ac.jp}}  
\vskip 0.5cm
 
 $^{\dagger}$Institute of Physics, University of Tokyo, Komaba, Tokyo, 153-8902 Japan  \\
 $^{\ast}$Institute for Cosmic Ray Research, University of Tokyo, Tanashi, Tokyo, 188-8502 Japan
\end{center}

\vskip 1cm
\begin{center} 
\begin{bf}
Abstract
\end{bf}
\end{center}
We study delocalization transition in a one-dimensional electron 
system with quenched disorder by using supersymmetric (SUSY) methods.
Especially we focus on effects of nonlocal correlation of disorder,
for most of studies given so far considered $\delta$-function type
white noise disorder.
We obtain wave function of the ``lowest-energy" state which dominates
partition function in the limit of large system size.
Density of states is calculated in the scaling region.
The result shows that delocalization transition is stable against
nonlocal short-ranged correlation of disorder.
Especially states near the band center are enhanced by the correlation
of disorder which partially suppresses random fluctuation of disorder.
Physical picture of the localization and the delocalization transition is discussed.

\newpage
\setcounter{footnote}{0}

\section{Introduction}
Quenched disorder plays important roles in various physical phenomena.
Anderson localization is one of these examples\cite{Anderson}.
Especially  in two and lower dimensions almost all states are localized,
and extended delocalized states are isolated points in physical parameter
region.
The transition between quantum Hall plateaus is such an example.

For the transition between integer quantum Hall plateaus,
useful model, named network model, has been proposed\cite{NM}, which incorporates
effects of localization and quantum tunneling in a strong magnetic field.
Numerical studies indicate that the network model belongs to the same
universality class of the transition between integer quantum Hall plateaus.
However it is rather difficult to solve the network model analytically because
there exists no controllable parameter for perturbative expansion and also
because of nature of quenched disorder itself.
Compared with the two-dimensional (2D) systems, one dimensional (1D)
systems are more tractable.
These 1D systems include Dyson's study on random strength harmonic
springs\cite{Dyson}, random Ising model by McCoy and Wu\cite{MW}, 
and more recently random exchange spin chains\cite{DF}
and random hopping tight-binding (RHTB) model\cite{CDM,BF,M}.

Recently supersymmetric (SUSY) methods appear useful for handling the 
quenched disorder.
SUSY methods are applied to the network model\cite{KM}, the 1D RHTB model\cite{BF} etc.
In this paper we shall revisit the 1D RHTB model by
applying the SUSY methods.
Especially we shall consider {\em nonlocally correlated} quenched disorder
and study stability of the delocalization transition which exists in the case
of the $\delta$-function-type white noise disorder. 
The model contains two parameters which control magnitude
of fluctuation and correlation length of disorder.
We expect that we can get detailed physical picture of (de)localization transition
from calculations of density of states, Green's functions, etc. as a function
of the above parameters.
Another motivation of the present work is rather technical, i.e., we show
how to use the SUSY methods for nonlocally correlated disorder systems.

Hamiltonian of the RHTB model is given by
\begin{equation}
{\cal H}_{RHTB}=-\sum_n t_n\Big(c^{\dagger}_nc_{n+1}+c^{\dagger}_{n+1}c_n\Big),
\label{RHTB}
\end{equation}
where $c_n$ is annihilation operator of spinless fermion at site $n$
and $t_n$'s are random hopping parameters.
As we shall see, staggered part of fluctuation of $t_n$'s plays an
important role at low energies and generates energy gap for most
of fermion modes.
Extended excitations are located near the band center, and they are
described by a Dirac fermion with randomly varying mass.

In the SUSY methods, bosonic variables are introduced.
In terms of them, density of states, Green's functions, etc are expressed 
in a compact way.
Average over random variables of {\em nonlocal} correlation can be performed
by introducing another bosonic variable, and because of that there appear 
couplings between the fermion and its bosonic SUSY partner.
The model is reduced to a quantum mechanical system of one fermionic and 
two bosonic variables.
The lowest-energy state which dominates the partition function is annihilated
by SUSY charges.
We obtain wave function of the lowest-energy state and calculate density
of states near the band center.

This paper is organized as follows;
In Sect.2, we shall review the RHTB model, its continuum limit
and SUSY methods which are applied for nonlocally correlated disorder.
Most of our notations follow those by Balents and Fisher in Ref.\cite{BF}.
In Sect.3, transfer Hamiltonian is obtained regarding the spatial
coordinate as time.
Then the system is reduced to a quantum mechanical system.
In Sect.4, we shall obtain an equation of motion of the ground state,
which dominates the partition function and density of states.
Section 5 is devoted for solution to equation of motion, and in Sect.6 
by using the ground state solution we shall calculate density of states
of fermions.
Section 7 is devoted for discussion.

In this paper, we calculate density of states of fermions.
We shall report Green's functions in a future publication\cite{IK}.
These calculations are quite useful for understanding localization
and the delocalization transition.

\setcounter{equation}{0}

\section{Continuum limit and SUSY}

The random hopping parameters fluctuate as 
\begin{equation}
t_n = t+\delta t_n,
\label{tn}
\end{equation}
where $t$ is some finite constant and we normalize as $t={1 \over 2a_l}$
and $a_l$ is the lattice spacing.
We assume that r.m.s. of random variables $\delta t_n$ is sufficiently small
compared with the band width $t$.
Continuum limit of the system (\ref{RHTB}) is then easily obtained.
We shall focus on modes near the band center because delocalization
transition occurs there.
Excitations near the band center of fermions are described by smoothly varying 
right and left-moving fermion fields $\psi_R$ and $\psi_L$,
\begin{equation}
c_n =e^{ik_Fn}\psi_R(x)+e^{-ik_Fn}\psi_L(x), \; \; k_F={\pi \over 2},
\label{fermionRL}
\end{equation}
where continuous coordinate $x=na_l$.

The most relevant part of the random hopping parameters $\delta t_n$
is their staggered part, $\delta t_n \sim (-1)^n m(x)$.
Then in terms of $\psi_R(x)$ and $\psi_L(x)$, we obtain continuum limit of the 
Hamiltonian (\ref{RHTB}) as 
\begin{eqnarray}
{\cal H}_c &=& -\int \!\!\mbox{dx} \; \Big[\psi_R^{\dagger}i\partial_x\psi_R-
\psi_L^{\dagger}i\partial_x\psi_L  
 -im(x)(\psi_R^{\dagger}\psi_L-\psi_L^{\dagger}\psi_R)  \Big],  \nonumber   \\
 &=&-\int \!\!\mbox{dx} \; \psi^{\dagger}h \psi,   \nonumber   \\
 h &=& -i\sigma^z\partial_x+m(x)\sigma^y,  \;\; \psi=(\psi_R\psi_L)^t.
\label{Hcont}
\end{eqnarray}
 The random mass $m(x)$ is decomposed into a uniform and random piece as
\begin{equation}
m(x)=m_0+\phi(x),
\label{random} 
 \end{equation}
 where $[\phi]_{ens}=0$ and 
\begin{equation}
[\phi(x)\phi(y)]_{ens}={g \over 2\lambda}\exp (-|x-y|/\lambda),
\label{phi}
\end{equation}
with positive parameters $g$ and $\lambda$. 
It is easily verified 
$$
\int \!\!\mbox{dx} \; [\phi(x)\phi(y)]_{ens}=g,
$$ 
and 
$$[\phi(x)\phi(y)]_{ens}
\rightarrow g\delta(x-y)\; \;  \mbox{as} \;\;\lambda \rightarrow 0.
 $$
It is obvious that the parameter $g$ controls magnitude of fluctuation
and $\lambda$ is the correlation length of disorder. 

Single fermion Green's function at energy $\omega$ is defined as 
\begin{equation}
G_{\alpha\beta}(x,y; i\omega)=\langle x,\alpha|{1 \over h-i\omega}|y,\beta
\rangle , \;\; \alpha, \beta=R,L,
\label{Green1}
\end{equation}
where 
$|x,\alpha \rangle$ is a {\em normalized} position eigenstate of fermion at $x$ and 
chirality $\alpha$.
By functional integral,
\begin{eqnarray}
G_{\alpha\beta}(x,y; i\omega)&=&i\langle \psi_{\alpha}(x)\bar{\psi}_{\beta}(y)
\rangle_{\psi}   \nonumber   \\
&=& {1 \over Z_{\psi}}\int {\cal D}\psi {\cal D}
\bar{\psi}\psi_{\alpha}(x)\bar{\psi}_{\beta}(y)e^{-S_{\psi}},  \nonumber \\
S_{\psi}&=& \int \!\!\mbox{dx} \; \bar{\psi}(ih +\omega )\psi,  \nonumber  \\
Z_{\psi}&=&\int {\cal D}\psi {\cal D}\bar{\psi}e^{-S_{\psi}}.
\label{Fintergal}
\end{eqnarray}
Ensemble averaged Green's function is obtained from (\ref{Green1}) as 
\begin{equation}
\overline{G}_{\alpha\beta}(x-y; i\omega)=\Big[\langle x,\alpha|{1 \over h-i\omega}|y,\beta
\rangle \Big]_{ens},
\label{Green2}
\end{equation}
where the ensemble average is taken with respect to $\phi(x)$ in (\ref{Hcont})
and (\ref{random}) according to (\ref{phi}).
To take the ensemble average seems formidable.
However by introducing SUSY partner, Green's function 
$\overline{G}_{\alpha\beta}(x-y; i\omega)$
etc can be expressed by a functional integral in a compact way.
Point is that integration over the bosonic SUSY partner cancels the fermionic
determinant and normalization of the fermionic states is automatically taken into
account.
Introducing the bosonic superpartner $\xi$,
\begin{equation}
\overline{G}_{\alpha\beta}(x-y; i\omega)=i\langle \psi_{\alpha}(x)\bar{\psi}_{\beta}(y)
\rangle_S,
\label{GSS}
\end{equation}
where
\begin{equation}
\langle {\cal A}\rangle_S=\Bigg[ \int {\cal D}\psi {\cal D}
\bar{\psi}{\cal D}\xi {\cal D}\bar{\xi} {\cal A} \;
e^{-S}\Bigg]_{ens},
\label{AS}
\end{equation}
with
\begin{equation}
S=\int\!\!\mbox{dx} \; \Big[\bar{\psi}(ih+\omega)\psi+\bar{\xi}(ih+\omega)\xi\Big].
\label{SS}
\end{equation}

The ensemble average in (\ref{AS}) can be performed in the following way.
We first notice identity such as
\begin{equation}
(-\lambda^2{\partial_x}^2 + 1)\frac{1}{2\lambda}\exp(-|x-y|/\lambda)
=\delta(x-y).
\label{greenphi}
\end{equation}
Then the ensemble average can be converted into the functional integral form,
\begin{eqnarray}
[\phi(x)\phi(y)...]_{ens}&=&\int {\cal D}\phi(x') 
(\phi(x)\phi(y)...)
\exp(-S_\phi[\phi(x')]), \label{pathphi} \\
S_\phi[\phi(x)]&=&\int \!\!\mbox{dx}\,\, 
\;
\frac{1}{4g}\phi(x)(-\lambda^2{\partial_x}^2 + 1)\phi(x).
\label{Sphi}
\end{eqnarray}
From (\ref{AS}), (\ref{SS}) and (\ref{pathphi}), the expectation value of 
operator ${\cal A}$ is given by
\begin{equation}
\langle{\cal A}\rangle_S=\int 
{\cal D}\psi{\cal D}\bar{\psi}{\cal D}\xi{\cal D}\bar{\xi}
{\cal D}\phi\mbox{ } {\cal A}\; \exp(-(S+S_\phi)).
\label{AS2}
\end{equation}
The above total action is obviously SUSY under $\psi \leftrightarrow \xi$
and has such a form that the SUSY partners $\psi$ and $\xi$ couple to
the ``dynamical" real scalar field $\phi$ which is ``singlet" under SUSY
transformation.

In the following section, we shall obtain a transfer Hamiltonian by regarding
the spatial coordinate $x$ as time.
The system reduces to a quantum mechanical system with the two bosonic
and one fermionic variables.

\setcounter{equation}{0}
\section{Transfer Hamiltonian}
In this section, we shall obtain a transfer Hamiltonian by regarding the spatial
coordinate $x$ as time in the functional integral representation (\ref{AS2}).
Then the system reduces to a quantum mechanical system.
In Ref.\cite{BF}, the following canonical creation and annihilation operators
are introduced corresponding to the functional integral variables,
\begin{eqnarray}
 \psi_R  \rightarrow F_{\uparrow}, \; \; \;
\bar{\psi}_R \rightarrow F^{\dagger}_{\uparrow},  &&
 \psi_L \rightarrow -F_{\downarrow}, \;\; \;
\bar{\psi}_L \rightarrow F^{\dagger}_{\downarrow}, \nonumber  \\
 \xi_R \rightarrow B_{\uparrow}, \;\;\; 
 \bar{\xi}_R \rightarrow B^{\dagger}_{\uparrow}, &&  
 \xi_L \rightarrow -B_{\downarrow}\;\;\;
 \bar{\xi}_L \rightarrow -B^{\dagger}_{\downarrow}.
\label{oper}
\end{eqnarray}
It is useful to define fermionic and bosonic spin operators,
\begin{eqnarray}
&& \vec{{\cal J}}_f={1 \over 2}F^{\dagger}\vec{\sigma}F,  \nonumber \\
&& \vec{{\cal J}}_b={1 \over 2}\bar{B}\vec{\sigma}B,
\label{spin}
\end{eqnarray}
where
\begin{equation}
\bar{B}=B^{\dagger}\sigma^z.
\label{Bbar}
\end{equation}
It is proved that $\vec{{\cal J}}_f$ and $\vec{{\cal J}}_b$ satisfy $SU(2)$ and 
$SU(1,1)$ algebras, respectively, and they commute with SUSY charges
$Q$ and $\bar{Q}$,
\begin{equation}
Q=\bar{B}F, \; \; \bar{Q}=F^{\dagger}B.
\label{SUSYc}
\end{equation}
Transfer Hamiltonian of $F_{\sigma}$ and $B_{\sigma}$ 
$(\sigma=\uparrow, \downarrow)$ part is written
in terms of the spin operators.

Transfer Hamiltonian of $\phi$ is also obtained from (\ref{Sphi}).
The system of $\phi$ is nothing but a simple harmonic oscillator
linearly coupled with the SUSY spin ${\cal J}={\cal J}_f+{\cal J}_b$.
In terms of the spin operators and the canonical boson operators of the 
harmonic oscillator $a, \; a^{\dagger}$ which correspond to $\phi$,
Hamiltonian of the system is given as,
\begin{equation} 
H=2\omega{\cal J}^{z}+2m_0{\cal J}^{x}+\sqrt{\frac{4g}{\lambda}}{\cal J}^{x}(a+a^\dagger)+\frac{1}{\lambda}(a^\dagger a+\frac{1}{2}).
\label{Hamiltonian0}
\end{equation}

Fermionic states of $F_{\sigma}$ are specified by 
representations of $SU(2)$.
Similarly bosonic states of $B_{\sigma}$ form multiplet of irreducible 
representations of $SU(1,1)$, which are specified by total spin 
 \begin{equation}
 J^2=(N_B^2+2N_B)/4, \;\; N_B=\bar{B}B=B^{\dagger}_{\uparrow}B_{\uparrow}
 -B^{\dagger}_{\downarrow}B_{\downarrow},
 \label{totalJ}
 \end{equation}
and $z$-component of spin $J^z$, i.e., 
\begin{eqnarray}
J^2|jn\rangle &=& j(j+1)|jn\rangle, \nonumber  \\
J^z|jn\rangle &=& \left[ {1+|2j+1| \over 2}+n\right] |jn\rangle,
\label{SU11}
\end{eqnarray}
 where it should be remarked here that the total
spin takes half integers, namely, $j=0,\pm 1/2, \pm 1,\cdots$.
 
 In Ref.\cite{BF}, structure of the above quantum mechanical states of the SUSY
 partners is studied in detail and it is shown that
 SUSY-invarinat states are explicitly given as follows;
 \begin{eqnarray}
 |n\rangle_0 =\cases{{1 \over \sqrt{2}}[|-1/2,n\rangle \otimes |\downarrow\rangle_F
 +|-1/2, n-1\rangle\otimes |\uparrow\rangle_F], & $n > 0$ \cr
                            |-1/2,0\rangle\otimes|\downarrow\rangle_F, & $n=0$ \cr }
\label{SUSYinv}                     
\end{eqnarray}                      
where $|\sigma\rangle_F$ is the fermionic state of spin $\sigma$.

\setcounter{equation}{0}
\section{The ground state}

From the functional-integral representation (\ref{AS2}),
it is obvious that the partition function 
$Z=\mbox{Tr}((-1)^{N_f}e^{-LH})$, where $N_f=$ fermion number
and $L=$ system size, acquires contribution 
only from the scalar field $\phi$ with the action $S_{\phi}$
because fermionic and bosonic determinants cancel with each other
for the SUSY partners.
Then in the large system size limit $L\rightarrow \infty$,
$Z=e^{-L/2\lambda}$.
The ``ground state" which satisfies 
\begin{equation}
H|0\rangle={1 \over 2 \lambda}|0\rangle, 
\label{groundeq0}
\end{equation}
or
\begin{equation}
\Big(2\omega{\cal J}^{z}+2m_0{\cal J}^{x}+\sqrt{\frac{2}{\lambda}}{\cal J}^{x}(a+a^\dagger)+\frac{1}{\lambda}a^\dagger a\Big)|0\rangle =0,
\label{ground}
\end{equation}
dominates the partition function and the density of states of fermions
which we shall calculate in this paper.
In Eq.(\ref{ground}), we have rescaled the parameters as 
$\omega \rightarrow 2g\omega,\; m_0 \rightarrow 2gm_0$ and $\lambda
\rightarrow {\lambda \over 2g}$.
This rescaling leads the general Hamiltonian to the one of $g={1\over 2}$.
After calculation, we shall scale back the above parameters
and recover dependence on $g$.

The SUSY vacuum should be invariant under SUSY
transformation and it must be annihilated by both $Q$ and 
$\bar{Q}$, and therefore it is given by a sum of a direct product
of $|n\rangle_0$ in Eq.(\ref{SUSYinv}) and $|m\rangle_H$, which is the 
eigenstates of the harmonic oscillator, i.e. ;
\begin{equation} \label{e22}
        |0\rangle=\sum_{n,m}\phi_{n,m}|n\rangle_0|m\rangle_H,
\label{vacuum0} 
\end{equation}
where $|m\rangle_H={1 \over \sqrt{m!}}(a^{\dagger})^m|0\rangle_H$
with the ``vacuum" of the harmonic oscillator $|0\rangle_H$.
Norm of the state (\ref{vacuum0}) is given by\footnote{As shown in Ref\cite{BF},
right and left states must be distinguished in the present system because
of the non-hermiticy of the Hamiltonian.
Then only $\phi_{n=0,m}$ contributes to the norm.}
\begin{equation}
\langle 0|0\rangle =\sum _m |\phi_{0,m}|^2.
\label{norm}
\end{equation}

Let us insert (\ref{vacuum0}) into (\ref{ground}),  
\begin{eqnarray} \label{e23}
    && \sum_{n,m}\phi_{n,m}\biggl(2\omega n|n\rangle_0|m\rangle_H
+ m_0\bigl((n+1)|n+1\rangle_0 - (n-1)|n-1\rangle_0\bigr)
|m\rangle_H \nonumber\\
&+& \sqrt{\frac{1}{2\lambda}}\bigl((n+1)|n+1\rangle_0 - (n-1)|n-1\rangle_0\bigr)(\sqrt{m+1}|m+1\rangle_H+\sqrt{m}|m-1\rangle_H) \nonumber \\
&+& \frac{1}{\lambda}m |n\rangle_0|m\rangle_H\biggr)=0.
\end{eqnarray}
Rearranging the indecies $n$ and $m$ of the subscripts of $\phi_{n,m}$
and the labels of states, we obtain equation of $\phi_{n,m}$
for the vacuum,
\begin{eqnarray} \label{e24}
       && (2\omega n + \frac{1}{\lambda}m)\phi_{n,m}
+ m_0\cdot n\cdot (\phi_{n-1,m}-\phi_{n+1,m}) \nonumber \\
&&+ \sqrt{\frac{1}{2\lambda}}n(\sqrt{m+1}(\phi_{n-1,m+1}-\phi_{n+1,m+1})
+ \sqrt{m}(\phi_{n-1,m-1}-\phi_{n+1,m-1}))=0.\nonumber \\
\end{eqnarray}
Equation (\ref{e24}) for $n=0$ can be solved
easily
and the solution is
\begin{eqnarray} \label{e24a}
        \phi_{0,m}=\left\{
\begin{array}{ll}
        \displaystyle{0,}& \qquad m\ne 0,\\
        \displaystyle{1,}& \qquad m= 0,\\ 
\end{array}\right. 
\end{eqnarray}
where the normalization comes from the condition
$\langle 0|0\rangle=1$.

By solving Eq.(\ref{e24}), we can calculate the density of states
of fermions.
However, it is not easy to obtain solution for arbitrary $\lambda$.
Then in the following section, we shall solve Eq.(\ref{e24}) 
by perturbative calculation in powers of $\lambda$, for $\lambda=0$
corresponds to the case of $\delta$-function type white noise.


\setcounter{equation}{0}

\section{Solution}

In this section we shall obtain wave function of the vacuum state
by solving (\ref{ground}).
We consider the case in which $\lambda$ is small.
The limit $\lambda \rightarrow 0$ corresponds to the short-range limit
which is discussed in Ref.\cite{BF}.

In order to see the correspondence between the transfer Hamiltonian
of the present system (\ref{Hamiltonian0}) and that of the short-range limit 
studied in Ref.\cite{BF},
\begin{equation}\label{p1}
H_0=2\omega{\cal J}^z + 2m_0{\cal J}^x - 2({\cal J}^x)^2,
\end{equation}
we rewrite the Hamiltonian (\ref{Hamiltonian0}) as follows,
\begin{equation}\label{p2}
H=H_0+\frac{1}{\lambda}(a^\dagger + \sqrt{2\lambda}{\cal J}^x)
(a + \sqrt{2\lambda}{\cal J}^x)+\frac{1}{2\lambda}.
\end{equation}
From (\ref{p1}), it is easily seen that the second term in Eq.(\ref{p2}) 
should vanish in the $\lambda\rightarrow 0$.

The ground state $|0\rangle$ of the transfer Hamiltonian (\ref{Hamiltonian0}) will 
be obtain by the perturbative calculation from
the state $|0\rangle^{(0)}$, which is annihilated by the operator
$(a + \sqrt{2\lambda}{\cal J}^x)$,
\begin{equation}\label{p3}
(a + \sqrt{2\lambda}{\cal J}^x)|0\rangle^{(0)}=0.
\end{equation}
Decomposing the state $|0\rangle^{(0)}$ into the states
$|n\rangle_0 |m\rangle_H$, namely, 
$|0\rangle^{(0)}=\sum \phi^{(0)}_{n,m}|n\rangle_0 |m\rangle_H$,
Eq.(\ref{p3}) leads to equations of 
$\phi^{(0)}_{n,m}$,
\begin{equation}\label{p4}
\sqrt{m+1}\phi^{(0)}_{n,m+1}+
\sqrt{\frac{\lambda}{2}}n(\phi^{(0)}_{n-1,m}-\phi^{(0)}_{n+1,m})=0.
\end{equation}
Equation (\ref{p4}) shows that, in the limit $\lambda\rightarrow 0$,
$\phi^{(0)}_{n,m}=0$ for $m\geq 1$ and therefore $\phi^{(0)}_{n,0}=\phi_{n}$, 
where $\phi_n$ is the wave function for $\lambda \rightarrow 0$ 
defined in Ref.\cite{BF}, i.e., 
\begin{equation}
H_0\sum_n \phi_n|n\rangle_0=0.
\label{H0sol}
\end{equation}

We shall consider higher-order corrections of
 finite $\lambda$. 
Equation (\ref{p4}) can be easily solved as
\begin{equation}\label{p5}
\phi^{(0)}_{n,m}=\frac{1}{\sqrt{m!}}(2\lambda)^{\frac{m}{2}}
(n\Delta_n)^{m}\phi^{(0)}_{n,0},
\end{equation}
where $\Delta_n$ is the difference operator with respects 
to the index $n$, i.e., 
$\Delta_n\phi^{(0)}_{n,m}=\frac{\phi^{(0)}_{n+1,m}-\phi^{(0)}_{n-1,m}}{2}$,
and this tells us the order of $\phi^{(0)}_{n,m}$;
\begin{equation}\label{p5.5}
\phi^{(0)}_{n,m}\sim\lambda^{\frac{m}{2}}.
\end{equation}
Now let us turn to the estimation of  the difference between the states 
$|0\rangle$ and $|0\rangle^{(0)}$. We decompose the former
as
\begin{equation}\label{p6}
|0\rangle=\sum_{n,m}(\phi^{(0)}_{n,m}+\delta\phi_{n,m})
|n\rangle_0 |m\rangle_H,
\end{equation}
where $\delta\phi_{n,m}$ comes from difference 
between $|0\rangle$ and $|0\rangle^{(0)}$. 
Substituting (\ref{p6}) into (\ref{groundeq0}) we obtain relation between
$\phi^{(0)}_{n,m}$ and $\delta\phi_{n,m}$,
\begin{equation}\label{p7}
\sum_{n,m}(H_0\phi^{(0)}_{n,m}+H_{\lambda}\delta\phi_{n,m})
|n\rangle_0 |m\rangle_H=0,
\end{equation}
where $H_{\lambda}\equiv H-\frac{1}{2\lambda}$. 
Noting $H_\lambda|0\rangle^{(0)}=H_0|0\rangle^{(0)}$,
we rewrite (\ref{p7}) as follows,
\begin{eqnarray}\label{p8}
&&\sum_{n,m}
\Big( H_0\phi^{(0)}_{n,m}+(H_0+2({\cal J}^x)^2+\frac{m}{\lambda})
\delta\phi_{n,m} \nonumber \\
&&+\sqrt{\frac{2}{\lambda}}{\cal J}^x
(\sqrt{m}\delta\phi_{n,m-1}+\sqrt{m+1}\delta\phi_{n,m+1})
\Big)|n\rangle_0 |m\rangle_H=0.
\end{eqnarray}
For each $m$, the terms should be the same order of 
$\lambda$, i.e.,
\begin{equation}\label{p8.5}
H_0\phi^{(0)}_{nm}
\sim\frac{1}{\lambda}\delta\phi_{n,m}
\sim\frac{1}{\sqrt{\lambda}}(\delta\phi_{n,m-1}+\delta\phi_{n,m+1}).
\end{equation}
From the fact that $H_0$ does not depend on $\lambda$ and (\ref{p5.5}), we obtain
\begin{equation}\label{p9}
\delta\phi_{n,m}\sim\lambda^{(\frac{m}{2}+1)},
\end{equation}
which is of higher order of $\lambda$ than $\phi^{(0)}_{n,m}$ for the 
same $m$. 
From the above result, we can estimate order of the term depending
on $\lambda$ in (\ref{p2}). 
From Eqs.(\ref{p5.5}), (\ref{p9}) and the following equation,
\begin{eqnarray}\label{p10}
&&(a^{\dagger}+\sqrt{2\lambda}{\cal J}^x)(a+\sqrt{2\lambda}{\cal J}^x)
|0\rangle \nonumber \\
&&=\sum_{n,m}\bigl(m\delta\phi_{n,m}+\sqrt{2\lambda}n
\Delta_n(\sqrt{m}\delta\phi_{n,m-1}+\sqrt{m+1}\delta\phi_{n,m+1})
\nonumber \\
&&\hspace{10 mm}+2\lambda(n\Delta_n)^2\delta\phi_{n,m}\bigr)
|n\rangle_0|m\rangle_H,
\end{eqnarray}
 we find that the above term is ${\cal O}(\lambda^{\frac{3}{2}}$).
Then it is clear that our estimation is consistent with the expectation
that in the limit $\lambda\rightarrow 0$ the present system becomes that
studied
in Ref.\cite{BF}. 

In order to calculate the density of states, we have to
solve Eq.(\ref{ground}). Decomposing it for each powers 
of $\lambda$ might be the easiest way to find  $\phi_{n,m}$. 
$\phi^{(0)}_{n,0}$ contains higher-order term of $\lambda$ 
than $\phi_n$. For $m=0$, Eq.(\ref{p8}) reads
\begin{equation}\label{w8}
\sum_n \left( H_0(\phi^{(0)}_{n,0}-\phi_n)+(H_0+2({\cal J}^x)^2)
\delta\phi_{n,0}+\sqrt{\frac{2}{\lambda}}{\cal J}^x \sqrt{m+1}
\delta\phi_{n,1}\right)|n\rangle_0 =0,
\end{equation}
where we use the fact that $H_0$ annihilates the vacuum 
for $\lambda=0$, and one can find that $\phi^{(0)}_{n,0}-\phi_n$ 
is ${\cal O}(\lambda)$.
Then from (\ref{p5}) we can see that $\phi^{(0)}_{n,m}$ contains 
a term of ${\cal O}(\lambda^{\frac{m}{2}+1})$ 
as the next-leading-order term, and therefore we can decompose it as 
$\phi^{(0)}_{n,m}=\hat{\phi}_{n,m}+\lambda\breve{\phi}_{n,m}$, 
and $\hat{\phi}_{n,m}$ and $\breve{\phi}_{n,m}$ are both 
${\cal O}(\lambda^{\frac{m}{2}})$.
We notice  the order of $\lambda$ in Eq.(\ref{p5}) and find
\begin{equation}
\hat{\phi}_{n,m}=\frac{1}{\sqrt{m!}}
(2\lambda)^{\frac{m}{2}}(n\Delta_n)^m\phi_n.
\label{hatphi}
\end{equation}
We insert the above $\phi^{(0)}_{n,m}$ into Eq.(\ref{p8}) and extract 
the terms in the order of ${\cal O}(\lambda^{\frac{m}{2}+1})$ 
and ${\cal O}(\lambda^{\frac{m}{2}})$;
\begin{eqnarray}\label{w9}
&&\sum_n \left(\lambda H_0\breve{\phi}_{n,m}
+(H_0+2({\cal J}^x)^2)\delta\phi_{n,m}
+\sqrt{\frac{2}{\lambda}}{\cal J}^x \sqrt{m+1}\delta\phi_{n,m+1}
\right)|n\rangle_0=0, \nonumber \\
&&\sum_n \left(H_0\hat{\phi}_{n,m}
+\frac{m}{\lambda}\delta\phi_{n,m}
+\sqrt{\frac{2}{\lambda}}{\cal J}^x \sqrt{m}\delta\phi_{n,m-1}
\right)|n\rangle_0=0.  
\end{eqnarray}
By inserting the second equation of (\ref{w9}) into the first one,
we obtain
\begin{eqnarray}\label{w10}
\sum_n H_0(\lambda\breve{\phi}_{n,m}+\delta{\phi}_{n,m})|n\rangle_0 
&&=\sqrt{\frac{2\lambda}{m+1}}\sum_n {\cal J}^x H_0\hat{\phi}_{n,m+1}
|n\rangle_0 \nonumber \\
&&=-\frac{2\lambda}{m+1}\sum_n {\cal J}^x H_0{\cal J}^x\hat{\phi}_{n,m}
|n\rangle_0,
\end{eqnarray}
where we have used Eq.(\ref{p4}) in ${\cal O}(\lambda^{\frac{m+1}{2}})$, 
namely, \[\sqrt{m+1}\sum\hat{\phi}_{n,m+1}|n\rangle_0+\sqrt{2\lambda}
{\cal J}^x\sum\hat{\phi}_{n,m}|n\rangle_0=0.\]
We shall solve Eq.(\ref{w10}) from now on.

Let us consider the $m=0$ case first, where $\hat{\phi}_{n,0}$ 
equals $\phi_n$. Appendix C in Ref.\cite{BF} gives us the solution
$\phi_n$.  
We review it here for completeness.
We recall that $\sum_n \phi_n|n\rangle_0$ is  
the vacuum state of the transfer Hamiltonian with $\lambda=0$, 
and thus $\phi_n$ satisfies the following equation, 
\begin{equation}\label{w11}
\left(2\omega n-2m_0n\Delta_n-2(n\Delta_n)^2\right)\phi_n=0.  
\end{equation}      
If $\phi(n,m_0)$ is a solution to Eq.(\ref{w11}),
then so is $(-1)^n\phi(n,-m_0)$, and thus general solution 
can be written as 
\begin{equation}\label{w12}
\phi_n=c_1\phi(n,m_0)+c_2(-1)^n\phi(n,-m_0),
\end{equation}
with constants $c_1$ and $c_2$.
Since  $\phi_n=1$ and $\phi_n=(-1)^n$ are 
solutions of the above equation for $\omega=m_0=0$, 
$\phi(n,m_0)$ for small $\omega$ and $m_0$ is a slowly
varying function of  $n$. 
This suggests that $\phi(n,m_0)$ is 
a ``continuum" function of $n$ and that enables us to replace 
discrete differences with derivatives. 
The differential equation for $\phi(n,m_0)$ is 
thus obtained as,
\begin{equation}\label{w13}
\left(2\omega n-2m_0 n\partial_n-2(n\partial_n)^2\right)
\phi(n,m_0)=0.
\end{equation}
In order to solve the above equation, we embed the integer $n$
into complex number and use the Laplace transformation,
\begin{equation}\label{w14}
\phi(n,m_0)=\int_{0}^{\infty}\!\!\mbox{dt}\,\,e^{-nt}\tilde{\phi}(t).
\end{equation}
By inserting the above Eq.(\ref{w14}) into Eq.(\ref{w13}), we have
\begin{equation}\label{w15}
\omega\tilde{\phi}(t)-(1-m_0)t\tilde{\phi}(t)
-t^2\partial_t\tilde{\phi}(t)=0,
\end{equation}
with the boundary condition $\tilde{\phi}(t)\rightarrow 0$
in the limit of $t\rightarrow 0$ or $t\rightarrow \infty$.
This boundary condition comes from the requirement that
$\phi(n,m_0)$ tends to vanish for large $n$.
We find the solution
\begin{equation}\label{w17}
\phi(n,m_0)=a_0\int_{0}^{\infty}\!\!\mbox{dt}\,\, t^{m_0-1}e^{-nt-\frac{\omega}{t}},
\end{equation}
where $a_0$ is a constant,
and the above solution can be written in terms of the modified Bessel function 
\begin{equation}\label{w18}
K_{\nu}(z)=K_{-\nu}(z)=\frac{1}{2}(\frac{z}{2})^{\nu}\int_{0}^{\infty}\!\!\mbox{dt}\,\, 
e^{-t-\frac{z^2}{4t}}t^{-\nu-1}
\end{equation}
as 
\begin{equation}\label{w19}
\phi(n,m_0)=
2a_0\Big(\frac{\omega}{n}\Big)^{\frac{m_{0}}{2}}K_{m_{0}}(2\sqrt{\omega n}).
\end{equation}
In order to obtain the constants $c_1 a_0$ and $c_2 a_0$, we use the boundary condition
\cite{BF},
\begin{eqnarray}
&& \phi_0 \simeq c_1\phi(1,m_0)+c_2\phi(1,-m_0)\simeq 1 \label{w19a},\\
&& \Delta_n \phi_1 \simeq c_1\partial_n\phi(1,m_0)-c_2\partial_n\phi(1,-m_0)\simeq 0.
\label{w19b}
\end{eqnarray}
Please note that the modified Bessel functions have a singularity at $n=0$. Eq.(\ref{w11}) is,
therefore, ill-defined for $n=0$, but we can expect the ``genuine" $\phi(0,m_0)$ in $\phi_0$
has almost the same value of $\phi(1,m_0)$, since the function $\phi(n,m_0)$ is assumed 
to be a slowly varying function.
For small $n$, we can approximate $\phi(n,m_0)$ as 
\begin{eqnarray}
&&\phi(n,m_0) \nonumber \\
&&=a_0\int^{\infty}_{0}\!\!\mbox{dt}\,\,t^{m_{0}-1}e^{-nt-\frac{\omega}{t}} \nonumber \\ 
&&\simeq \frac{a_0}{n^{m_{0}}} 
\bigl(\int^{\sqrt{n\omega}}_{0}\!\!\mbox{dt}\,\,t^{m_{0}-1}e^{-\frac{n\omega}{t}}
+\int^{\infty}_{\sqrt{n\omega}}\!\!\mbox{dt}\,\,t^{m_{0}-1}e^{-t} \bigr) \nonumber \\
&&\simeq \frac{a_0}{n^{m_{0}}}\bigl(\Gamma(m_{0})+(n\omega)^{m_{0}}\Gamma(-m_{0})\bigr)
\nonumber \\
&&\simeq \frac{a_0}{m_{0}} (n^{-m_{0}}-\omega^{m_{0}}), \nonumber \\
\end{eqnarray}
which coincides with the result obtained in the hard-wall approximation \cite{BF}.
Using this, Eqs.(\ref{w19a}) and (\ref{w19b}) give
\begin{eqnarray}
&&\frac{c_1 a_0}{m_{0}}(1-\omega^{m_{0}})+
\frac{c_2 a_0}{m_{0} \omega^{m_{0}}}(1-\omega^{m_{0}})=1, \\
&&c_1 a_0 = c_2 a_0.
\end{eqnarray}
Thus the constants are
\begin{equation}
c_1 a_0 = c_2 a_0 = \frac{m_{0} \omega^{m_{0}}}{1-\omega^{2m_{0}}}.
\end{equation}

Now let us return to Eq.(\ref{w10}). 
Renumbering the indecies of each terms, we have 
\begin{eqnarray}\label{w20}
&&(2\omega n-2m_{0}n\Delta_n-2(n\Delta_n)^2)
(\lambda\breve{\phi}_{n,m}+\delta{\phi}_{n,m})=\nonumber \\
&&=-\frac{2\lambda}{m+1}n\Delta_n(2\omega n-2m_{0}n\Delta_n-2(n\Delta_n)^2)
n\Delta_n\hat{\phi}_{n,m}.
\end{eqnarray}
Since $\hat{\phi}_{n,m}$ can be  decomposed into the form
\begin{equation}\label{w21.5}
\hat{\phi}_{n,m}=
c_1\phi(n,m,m_0)+c_2(-1)^{n+m}\phi(n,m,-m_0),
\end{equation}
we find that $\lambda\breve{\phi}_{n,m}+\delta{\phi}_{n,m}$ can be also
written as 
\begin{equation}\label{w21}
\lambda\breve{\phi}_{n,m}+\delta{\phi}_{n,m}=
\lambda(c_1\psi(n,m,m_0)+c_2(-1)^{n+m}\psi(n,m,-m_0)).
\end{equation}
As it is for $\phi_n$, we can assume that $\psi(n,m,m_0)$ is a slowly varying function
with respect to $n$. 
This again allows us to replace the difference operator $\Delta_n$ with 
the differential operator $\partial_n$.  
Eq.(\ref{w20}) can be rewritten in the simpler form;
\begin{eqnarray}\label{w22}
&&(2\omega n-2m_{0}n\partial_n-2(n\partial_n)^2)
(\psi(n,m,m_0)+\frac{2}{m+1}(n\partial_n)^2\phi(n,m,m_0)) \nonumber \\
&&=-\frac{2}{m+1} 2\omega n^2\partial_n\phi(n,m,m_0).
\end{eqnarray}
We explicitly solve the above equation for the case $m=0$. 
For $n\neq 0$,
\begin{eqnarray}\label{w23}
&&(2\omega -2(1+m_0)\partial_n-2n(\partial_n)^2)
(\psi(n,m_0)+2(n\partial_n)^2\phi(n,m_0))= \nonumber \\
&&=-4\omega n\partial_n\phi(n,m_0),
\end{eqnarray} 
where $\psi(n,m_0)=\psi(n,0,m_0)$. 
We again employ Laplace transformation to solve 
the above equation as before,
\begin{equation}\label{w24}
\psi(n,m_0)+2(n\partial_n)^2\phi(n,m_0)=
\int_{0}^{\infty}\!\!\mbox{dt}\,\,e^{-nt}\tilde{\psi}(t).
\end{equation}
From (\ref{w24}) and (\ref{w14}), Eq.(\ref{w23}) is rewritten as follows;
\begin{eqnarray}\label{w25}
(\omega-(1-m_{0})t-t^2\partial_t)\tilde{\psi}(t)&=&2\omega\partial_t(t\tilde{\phi}(t))
\nonumber \\
&=&2 a_0\omega\partial_t(t^{m_{0}}e^{-\frac{\omega}{t}}) \nonumber \\
&=&2 a_0\omega(m_{0} t^{m_{0} -1}+\omega t^{m_{0} -2})e^{-\frac{\omega}{t}}.
\end{eqnarray}
In general, we can put $\tilde{\psi}(t)=C(t)e^{-\frac{\omega}{t}}$.
Inserting this into Eq.(\ref{w25}), we find that $C(t)$ satisfies the following equation,
\begin{equation}\label{w26}
t^2\partial_t C(t)+(1-m_{0})tC(t)=-2a_0\omega(m_{0} t^{m_{0} -1}+\omega t^{m_{0} -2}).
\end{equation}
A special solution of this equation is 
$C(t)=2a_0\omega m_{0} t^{m_{0} -2}+a_0\omega^2 t^{m_{0} -3}$, and thus
we find
\begin{eqnarray}\label{w27}
&&\psi(n,m_0)=-2(n\partial_n)^2\phi(n,m_0)+
a_0\omega\int_{0}^{\infty}\!\!\mbox{dt}\,\,
(2m_{0} t^{m_{0} -2}+\omega t^{m_{0} -3})e^{-nt-\frac{\omega}{t}}\nonumber \\
&&=
-4a_0(n\partial_n)^2\bigl((\frac{\omega}{n})^{\frac{m_{0}}{2}}K_{m_{0}}(2\sqrt{n\omega})
\bigr)\nonumber \\
&&+2a_0\omega\bigl(2m_{0}(\frac{n}{\omega})^{\frac{1-m_{0}}{2}}K_{1-m_{0}}(2\sqrt{n\omega}) 
+\omega (\frac{n}{\omega})^{\frac{2-m_{0}}{2}}K_{2-m_{0}}(2\sqrt{n\omega})\bigr).
\end{eqnarray}
Then we find that the general solution to Eq.(\ref{w20}) for $m=0$ is given by
\begin{eqnarray}\label{w28}
&&\lambda\breve{\phi}_{n,0}+\delta{\phi}_{n,0}=f\lambda\phi_n
-2\lambda(n\Delta_n)^{2}\phi_n\nonumber \\ 
&&+2\lambda\omega \bigl(c_1 a_0 (2m_{0}(\frac{n}{\omega})^{\frac{1-m_{0}}{2}}
K_{1-m_{0}}(2\sqrt{n\omega}) 
+\omega (\frac{n}{\omega})^{\frac{2-m_{0}}{2}}K_{2-m_{0}}(2\sqrt{n\omega}))
\nonumber \\ 
&&+c_2 a_0 (-1)^n (-2m_{0}(\frac{n}{\omega})^{\frac{1+m_{0}}{2}}
K_{1+m_{0}}(2\sqrt{n\omega}) 
+\omega (\frac{n}{\omega})^{\frac{2+m_{0}}{2}}K_{2+m_{0}}(2\sqrt{n\omega}))\bigr),
\nonumber \\ 
\end{eqnarray}
where $f$ is a constant which is to be determined by the boundary 
condition Eq.(\ref{e24a}). 
For regular $\phi_n$'s, 
$(n\Delta_n)^{m}\phi_n=0$ for $n=0$, since this term is proportional
to $n$. This gives $\hat{\phi}_{0,m}=0$ for $m>0$, which can be  
verified by practical calculation. 
Using $\phi_0=1$, the condition Eq.(\ref{e24a}) requires
\begin{equation}\label{w29}
\lambda\breve{\phi}_{0,m}+\delta\phi_{0,m}=0,
\end{equation}
for arbitrary $m$.
We insert (\ref{w28}) into (\ref{w29}) for the case $m=0$ in order to
determine the constant $f$. For small $n$, the modified Bessel function
is approximated as
\begin{eqnarray}\label{w30}
&& 2(\frac{n}{\omega})^{\frac{1\pm m_{0}}{2}}K_{1\pm m_{0}}(2\sqrt{n\omega})
\simeq \omega^{\mp m_{0}-1}, \\
&& 2(\frac{n}{\omega})^{\frac{2\pm m_{0}}{2}}K_{2\pm m_{0}}(2\sqrt{n\omega})
\simeq \omega^{\mp m_{0}-2}. 
\end{eqnarray}
Then Eq.(\ref{w29}) for $m=0$ is 
\begin{eqnarray}\label{w31}
&&\lambda\breve{\phi}_{0,0}+\delta\phi_{0,0} \nonumber \\
&&=f\lambda + \lambda \bigl(c_1 a_0 \omega(2m_{0}\omega^{m_{0}-1}+\omega\omega^{m_{0}-2})
+c_2 a_0 \omega(-2m_{0}\omega^{-m_{0}-1}+\omega\omega^{-m_{0}-2})\bigr) \nonumber \\
&&=f\lambda - \lambda 
\frac{m_{0}\omega^{m_{0}}}{1-\omega^{2m_{0}}}\bigl(2m_{0}\frac{1-\omega^{2m_{0}}}{\omega^{m_{0}}}
-\frac{1+\omega^{2m_{0}}}{\omega^{m_{0}}}\bigr) \nonumber \\
&&=f\lambda-2\lambda m_{0}^{2}+\lambda m_{0}\frac{1+\omega^{2m_{0}}}{1-\omega^{2m_{0}}}=0.
\end{eqnarray}
Thus 
\begin{equation}\label{w32}
\phi_{n,0}= \phi_n + \lambda \psi_n,
\end{equation}
where
\begin{eqnarray}\label{w33}
&&\psi_n=(2m_{0}^{2}-m_{0}\frac{1+\omega^{2m_{0}}}{1-\omega^{2m_{0}}})\phi_n
-2(n\Delta_n)^{2}\phi_n
\nonumber \\
&&+2\omega \frac{m_{0}\omega^{m_{0}}}{1-\omega^{2m_{0}}}
\biggl( 2m_{0}(\frac{n}{\omega})^{\frac{1-m_{0}}{2}}
K_{1-m_{0}}(2\sqrt{n\omega}) 
+\omega (\frac{n}{\omega})^{\frac{2-m_{0}}{2}}K_{2-m_{0}}(2\sqrt{n\omega})
\nonumber \\ 
&&+ (-1)^n (-2m_{0}(\frac{n}{\omega})^{\frac{1+m_{0}}{2}}
K_{1+m_{0}}(2\sqrt{n\omega}) 
+\omega (\frac{n}{\omega})^{\frac{2+m_{0}}{2}}K_{2+m_{0}}(2\sqrt{n\omega}))\biggr).
\nonumber \\ 
\end{eqnarray}
The other components $\phi_{n,m}$ for $m\geq 1$ can be calculated in a similar
way to the above. 
Calculations are summarized in Appendix A.

\setcounter{equation}{0}
\section{The density of states}
Now we can calculate the density of states. The density of states is obtained 
from 
$\overline{G}_{\alpha\beta}$ as
\begin{equation}
\rho(\epsilon)=\frac{1}{\pi}\lim_{\omega\rightarrow 0}\mbox{Im}\bar{\cal G}(0;\epsilon+i\omega),
\end{equation}
where the averaged Green's function $\bar{\cal G}(x;i\omega)$ is given by, 
\begin{equation}
\bar{\cal G}(x;i\omega)=i^x \sum_{\alpha}(-1)^{\alpha x}\overline{G}_{\alpha\alpha}
(x;i\omega), \nonumber
\end{equation}
with $\alpha=0(1)$ for $R(L)$.
As shown in Ref.\cite{BF}, 
\begin{equation}
\bar{\cal G}(0;i\omega)=i\sum_{n,m}(-1)^n |\phi_{n,m}|^2,
\end{equation}
and we shall calculate $\rho(\epsilon)$
by performing analytic continuation $i\omega\rightarrow \epsilon+i\omega$.

The low-order corrections of $\lambda$ in 
the averaged Green function is given as 
\begin{eqnarray}
&&\bar{\cal G}(0;i\omega)=i\sum_{n=0}^{\infty}(-1)^n \phi_{n,0}^2
+i\sum_{n=0}^{\infty}(-1)^n \phi_{n,1}^2 \cdots \nonumber \\
&&=i\sum_{n}(-1)^n \phi_{n}^2+\nonumber \\
&&\hspace{3 mm}+2i\sum_{n}(-1)^n \phi_{n}(\lambda\breve{\phi}_{n,0}+\delta\phi_{n,0})
+i\sum_{n}(-1)^n \hat{\phi}_{n,1}^2 \nonumber \\
&&\hspace{3 mm}+{\cal O}(\lambda^{2}), \nonumber \\
\end{eqnarray}
where each raw corresponds to the order ${\cal O}(1)$, ${\cal O}(\lambda)$, 
and higher.
Inserting the $\hat{\phi}_{n,m}$ and $\lambda\breve{\phi}_{n,m}+\delta\phi_{n,m}$,
and using the formula \cite{GRA}
\begin{eqnarray}
&&\int_{0}^{\infty}\!\!\mbox{dn}\,\,\bigl(2(\frac{n}{\omega})^{\frac{\mu}{2}}K_{\mu}(2\sqrt{n\omega})
\cdot 2(\frac{n}{\omega})^{\frac{\nu}{2}}K_{\nu}(2\sqrt{n\omega})\bigr) \nonumber \\
&&\hspace{3 mm}=\int_{0}^{\infty}\!\!\mbox{dz}\,\,\frac{z}{2\omega}\cdot
2(\frac{z}{2\omega})^{\mu}K_{\mu}(z) \cdot 
2(\frac{z}{2\omega})^{\nu}K_{\nu}(z) \nonumber \\
&&\hspace{3 mm}=4(\frac{1}{2\omega})^{\mu+\nu+1}\cdot\frac{2^{\mu+\nu-1}}{1+\mu+\nu}
\Gamma(1+\mu)\Gamma(1+\nu),   
\label{Bfunc}
\end{eqnarray}
we have               
\begin{eqnarray}
&&\hspace{-5 mm}\sum_{n}(-1)^n \phi_n^2 
\simeq 2c_1 c_2 \int_{0}^{\infty}\!\!\mbox{dn}\,\,\phi(n,m_0)\phi(n,-m_0) 
\nonumber \\
&&\hspace{14 mm}\simeq \frac{2 m_0^2 \omega^{2m_0}}{\omega(1-\omega^{2m_0})^{2}} \\
&&\hspace{-5 mm}\sum_{n}(-1)^n \phi_{n}(\lambda\breve{\phi}_{n,0}+\delta\phi_{n,0}) 
\simeq -\frac{\lambda m_0^2\omega^{2m_0}}{(1-\omega^{2m_0})^2}
\bigl(\frac{2m_0(1+\omega^{2m_0})}{\omega(1-\omega^{2m_0})}-\frac{2}{3\omega}\bigr) \\
&&\hspace{-5 mm}\sum_{n}(-1)^n \hat{\phi}_{n,1}^2 \simeq 
\frac{2\lambda m_0^2\omega^{2m_0}}{(1-\omega^{2m_0})^2}
\bigl(\frac{2m_0^2}{\omega}-\frac{2}{3\omega}\bigr). \\
\nonumber
\end{eqnarray}
Higher-order terms can be calculated in a similar way. We are interested in
the averaged Green's function in the limit  $m_0\rightarrow 0$, since 
the delocalization transition occurs there. 
Collecting the higher-order terms, 
\begin{eqnarray}
\bar{\cal G}(0;i\omega)&&= \frac{i}{2\omega(\ln\omega)^{2}}
\nonumber \\
&& \hspace{3 mm}
+\lambda(\frac{i}{\omega(\ln\omega)^{3}})
\nonumber \\
&& \hspace{3 mm}
+\lambda^2(\frac{13i}{30\omega(\ln\omega)^{2}}+\frac{i}{3\omega(\ln\omega)^{3}}
+\frac{i}{2\omega(\ln\omega)^{4}})
\nonumber \\
&& \hspace{3 mm}
-\lambda^3\frac{1}{210}\frac{i}{\omega(\ln\omega)^{2}} 
\nonumber \\
&& \hspace{3 mm}
+{\cal O}(\lambda^4).
\label{barcalg}
\end{eqnarray}
Each term in the averaged Green's function has the form 
$\frac{i}{\omega(\ln \omega)^{\alpha}}$. We perform analytic continuation
in order to obtain the density of states $\rho(\epsilon)$ such as
\begin{eqnarray}
\mbox{Im}\frac{i}{\omega(\ln \omega)^{\alpha}}\rightarrow 
-\mbox{Im}\frac{1}{\epsilon(\ln \epsilon-i\frac{\pi}{2})^{\alpha}}
\simeq (-1)^{\alpha}\frac{\alpha\pi}{2\epsilon|\ln \epsilon|^{\alpha+1}},
\end{eqnarray} 
for small $\epsilon$.
Recovering the constant $g$, the density of states is obtained as follows,
\begin{eqnarray}
&&\hspace{-8 mm}\rho(\epsilon)
=\frac{1}{2\frac{\epsilon}{2g}\bigl(\ln\frac{\epsilon}{2g}\bigr)^3}\nonumber \\
&&-2g\lambda\frac{3}{2\frac{\epsilon}{2g}\bigl(\ln\frac{\epsilon}{2g}\bigr)^4}\nonumber \\
&&+4g^2\lambda^{2}\biggl(\frac{13}{30\frac{\epsilon}{2g}\bigl(\ln\frac{\epsilon}{2g}\bigr)^3}
-\frac{1}{2\frac{\epsilon}{2g}\bigl(\ln\frac{\epsilon}{2g}\bigr)^4}
+\frac{1}{\frac{\epsilon}{2g}\bigl(\ln\frac{\epsilon}{2g}\bigr)^5}\biggr) \nonumber \\
&&-8g^3\lambda^{3}\frac{1}{210\frac{\epsilon}{2g}\bigl(\ln\frac{\epsilon}{2g}\bigr)^3}\nonumber \\
&&+{\cal O}((g\lambda)^4).
\label{rho}
\end{eqnarray} 
The above expression (\ref{rho}) is the main result of this paper.

\setcounter{equation}{0}
\section{Discussion}

In the previous sections, we have obtained density of states of fermions
interacting with quenched disorder of nonlocal correlation.
The result is given by (\ref{rho}) and Figs.1 and 2.
It is obvious that unit of the energy $\epsilon$ is $g$, the magnitude of
disorder fluctuation, and $\rho(\epsilon)$ scales with $(g\lambda)$.
In the telegraph process of random disorder\cite{CDM}, ${g \over \lambda}$ and $\lambda$ 
correspond to the mean height and width of fluctuating mass $m(x)$, respectively.
Recently we studied the eigenvalue problem
\begin{equation}
h\psi=E\psi
\label{hpsi}
\end{equation}
by numerical calculation for telegraphic $m(x)$, and found that all the eigenvalues
are determined by the combination $(g\lambda)$ in unit $g$\cite{TTI}. 

In Fig.1, we show $\rho(\epsilon)$ for various $\lambda$ with a fixed  $g$.
It is easily seen that for $\epsilon$ smaller than $\epsilon_0\sim ge^{-{1 \over g\lambda}}$,
$\rho(\epsilon)$ is a increasing function of $\lambda$ whereas for $\epsilon>\epsilon_0$ 
it is a decreasing function.
That is, states near the band center are enhanced by the nonlocal correlation of disorder,
which partially suppresses random fluctuation of disorder.
From the above result, we expect that the number of extended states near the band center
is increased by the existence of  such kind of disorder correlation.

The above expectation can be verified by solving (\ref{hpsi}).
In order to study the field equation (\ref{hpsi}) in the multi-solitonic
background of $m(x)$, transfer-matrix formalism is very useful\cite{TTI}.
Using this method, we have obtained solutions to (\ref{hpsi})
in various configurations of $m(x)$.
We think that from these explicit form of solutions we can understand
characteristic properties of the present random system and their origins,
e.g., power-law decay of Green's functions, complex multi-fractal scaling\cite{H},
etc.

Results in \cite{TTI} are summarized as follows;
\begin{enumerate}
\item For quasi-periodic $m(x)$ almost all states are extended and
quasi-periodic.
Peaks of the wave functions appear quasi-periodically.
This is a reminiscence of Bloch's theorem although we are {\em not} imposing
the periodic boundary condition on the wave function.
This result remains correct even for $m_0=\langle m(x)\rangle \neq 0$.
\item For large random fluctuation of disorder close to the white noise,
almost all states tend to ``localized" even for small or vanishing $m_0$. 
They are divided into two categories; the first one simply has one peak 
or peaks close each other, whereas the second one has more than two peaks
which are spatially separated comparatively large distance.
The first one is obviously genuine localized state.
\item On suppressing randomness of disorder, the number of states of the second 
category increases.
\end{enumerate}
We think that the above results support our conclusion from the density
of states, i.e., the second category is main part of  extended states and 
the number of extended states is increased by nonlocal correlation
of disorder.
We also expect that on increasing correlation of disorder,
quasi-periodic states in quasi-periodic $m(x)$ come to contribute to 
physical quantities as extended state.

Sometimes  properties of the present disordered system
are explained by using the single zero-energy
wave function $\psi_{\pm} \propto e^{\pm \int^x \!\!\mbox{dx'}\,\,m(x')}$ \cite{ST}.  
However as the behavior of the density of states shows, a large number of low-energy
states are contributing to physical quantities.
Then it is important to see if these ``typical" low-energy modes have almost similar
behaviors to those of $\psi_{\pm}$.
Details will be reporteded in a future publication\cite{TTI}.
We also calculate Green's functions in the present system\cite{IK}.
All these results are quite useful for understanding localization and delocalization
transition.


\begin{center}
{\bf ACKNOWLEDGMENTS}
\end{center}
We are grateful to  K.Takeda and T.Tsurumaru for useful and enlightening
discussion.


\newpage
\appendix
\renewcommand{\theequation}{\Alph{section}.\arabic{equation}}
{\Large {\bf Appendix}}
\setcounter{equation}{0}
\section{Higher-order terms}
Higher-order corrections to the averaged Green's function are calculated  
similarly to the correction of ${\cal O}(\lambda)$. It is easy to see that
the limit, $m_0\rightarrow 0$, and the integration, Eq.(\ref{Bfunc}), are  
commutable with each other, and thus we take the limit $m_0\rightarrow 0$ first.

From the definition of $\hat{\phi}_{n,m}$, Eq.(\ref{hatphi}), and taking that limit,
we have
\begin{equation}
\hat{\phi}_{n,m}=-\frac{1}{2a_0\ln\omega}\phi(n,m,0)(1+(-1)^{n+m})
\;\;\;\;(m=1,2,3,\cdots), \nonumber 
\end{equation}
with
\begin{eqnarray}
\phi(n,1,0)&=&
-a_0\sqrt{2\lambda}\omega
\cdot2\bigl(\frac{n}{\omega}\bigr)^{\frac{1}{2}}
K_1(2\sqrt{n\omega}),
\nonumber \\
\phi(n,2,0)&=&\sqrt{2}a_0\lambda\biggl(-\omega\cdot
2\bigl(\frac{n}{\omega}\bigr)^{\frac{1}{2}}
K_1(2\sqrt{n\omega})+\omega^2\cdot
2\bigl(\frac{n}{\omega}\bigr)^{\frac{1}{2}}
K_1(2\sqrt{n\omega})\biggr),
\nonumber \\
\phi(n,3,0)&=&
-\frac{2}{\sqrt{3}}a_0\lambda^{\frac{3}{2}}
\biggl(\omega\cdot 2\bigl(\frac{n}{\omega}\bigr)^{\frac{1}{2}}
K_1(2\sqrt{n\omega})-3\omega^2\cdot
2\bigl(\frac{n}{\omega}\bigr)K_2(2\sqrt{n\omega}) \nonumber \\
&&\hspace{3 cm}
+\omega^{3}\cdot 2\bigl(\frac{n}{\omega}\bigr)^{\frac{3}{2}}
K_3(2\sqrt{n\omega})\biggr).\nonumber 
\end{eqnarray}

Solving Eq.(\ref{w20}), we have the higher-order terms in $\phi_{n,m}$
\begin{equation}
\lambda\breve{\phi}_{n,m}+\delta\phi_{n,m}
=-\frac{1}{2a_0\ln\omega}\psi(n,m,0)(1+(-1)^{n+m}) \;\;\;\;(m=1,2,3,\cdots),
\nonumber
\end{equation}
with
\begin{eqnarray}
\psi(n,1,0)&=&-\sqrt{2}a_0\lambda^{\frac{3}{2}}\biggl(
\omega\cdot 2\bigl(\frac{n}{\omega}\bigr)^{\frac{1}{2}}K_1(2\sqrt{n\omega})
-\frac{5}{2}\omega^2\cdot 2\bigl(\frac{n}{\omega}\bigr)
K_2(2\sqrt{n\omega})\nonumber \\
&&\hspace{3 cm}
+\frac{2}{3}\omega^3\cdot 2\bigl(\frac{n}{\omega}\bigr)^{\frac{3}{2}}
K_3(2\sqrt{n\omega}\biggr) \nonumber \\
\psi(n,2,0)&=&\frac{2\sqrt{2}}{3}a_0\biggl(\omega\cdot 2\bigl(\frac{n}{\omega}\bigr)^{\frac{1}{2}}
K_1(2\sqrt{n\omega})-\frac{13}{2}\omega^2\cdot 2\bigl(\frac{n}{\omega}\bigr)K_2(2\sqrt{n\omega})
\nonumber \\
&&\hspace{3 cm}+5\omega^3\cdot 2\bigl(\frac{n}{\omega}\bigr)^{\frac{3}{2}}K_3(2\sqrt{n\omega})
-\frac{3}{4}\omega^4\cdot 2\bigl(\frac{n}{\omega}\bigr)^{2}K_4(2\sqrt{n\omega})\biggr).
\nonumber 
\end{eqnarray}

From the above results, we have the higher corrections to the averaged Green's function;
Terms in ${\cal O}(\lambda)$ are
\begin{eqnarray}
&&\sum_n \hat{\phi}_{n,1}^2(-1)^n=-\frac{\lambda}{3}\frac{1}{\omega(\ln\omega)^2},
\nonumber \\
&& 2\sum_n \hat{\phi}_{n,0}(\lambda\breve{\phi}_{n,0}+\delta\phi_{n,0})(-1)^n
=\lambda\biggl(\frac{1}{3}\frac{1}{\omega(\ln \omega)^2}+\frac{1}{\omega(\ln\omega)^3}
\biggr), \nonumber 
\end{eqnarray}
 terms in ${\cal O}(\lambda^2)$ are
\begin{eqnarray}
&&\sum_n \hat{\phi}_{n,2}^2(-1)^n=\lambda^2\frac{2}{15}\frac{1}{\omega(\ln\omega)^2},
\nonumber \\
&& 2\sum_n \hat{\phi}_{n,1}(\lambda\breve{\phi}_{n,1}+\delta\phi_{n,1})(-1)^n
=\lambda^2\biggl(\frac{7}{30}\frac{1}{\omega(\ln \omega)^2}\biggr), 
\nonumber \\
&&\sum_n (\lambda\breve{\phi}_{n,0}+\delta\phi_{n,0})^2(-1)^n
=\lambda^2\biggl(\frac{1}{15}\frac{1}{\omega(\ln\omega)^2}+
\frac{1}{3}\frac{1}{\omega(\ln\omega)^3}+\frac{1}{2}\frac{1}{\omega(\ln\omega)^4}\biggr),
\nonumber 
\end{eqnarray}
and terms in ${\cal O}(\lambda^3)$ are
\begin{eqnarray}
&&\sum_n \hat{\phi}_{n,3}^2(-1)^n=-\lambda^3\frac{16}{315}\frac{1}{\omega(\ln\omega)^2},
\nonumber \\
&& 2\sum_n \hat{\phi}_{n,2}(\lambda\breve{\phi}_{n,2}+\delta\phi_{n,2})(-1)^n
=\lambda^3\biggl(\frac{31}{315}\frac{1}{\omega(\ln \omega)^2}\biggr), 
\nonumber \\
&&\sum_n (\lambda\breve{\phi}_{n,1}+\delta\phi_{n,1})^2(-1)^n
=-\lambda^3\biggl(\frac{11}{210}\frac{1}{\omega(\ln\omega)^2}\biggr).
\nonumber 
\end{eqnarray}
Combining these terms, we have the averaged Green function Eq.(\ref{barcalg}).


\newpage
{\bf Figure captions}\\

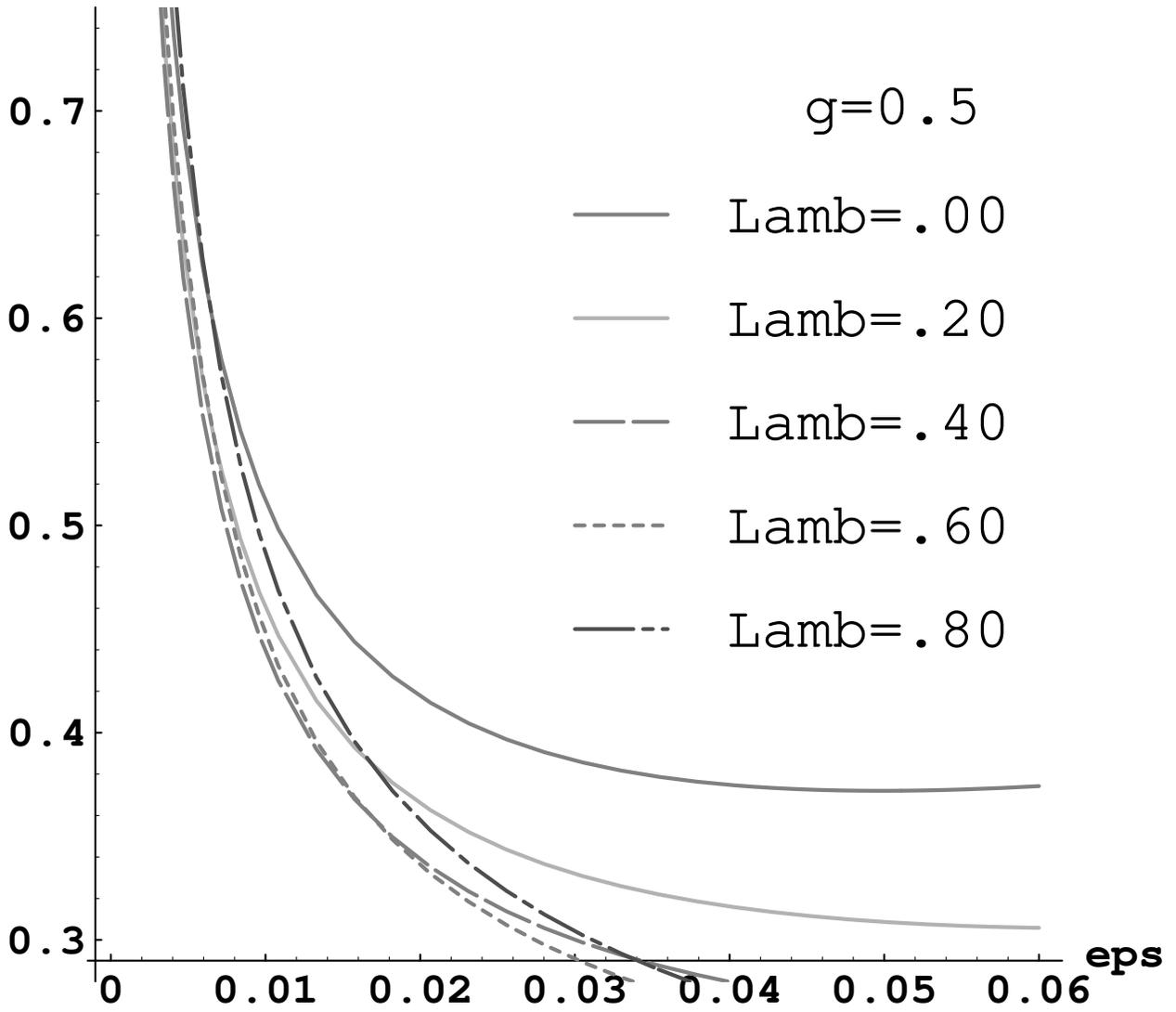
Fig. 1 : The density of states, as a function of the energy, 
whose $\lambda$ is varied from $0.00$ to $0.80$ at $g=0.5$.
The axis of abscissa stands for the energy ($\epsilon$) and 
the one of column for the density of states ($\rho(\epsilon)$).

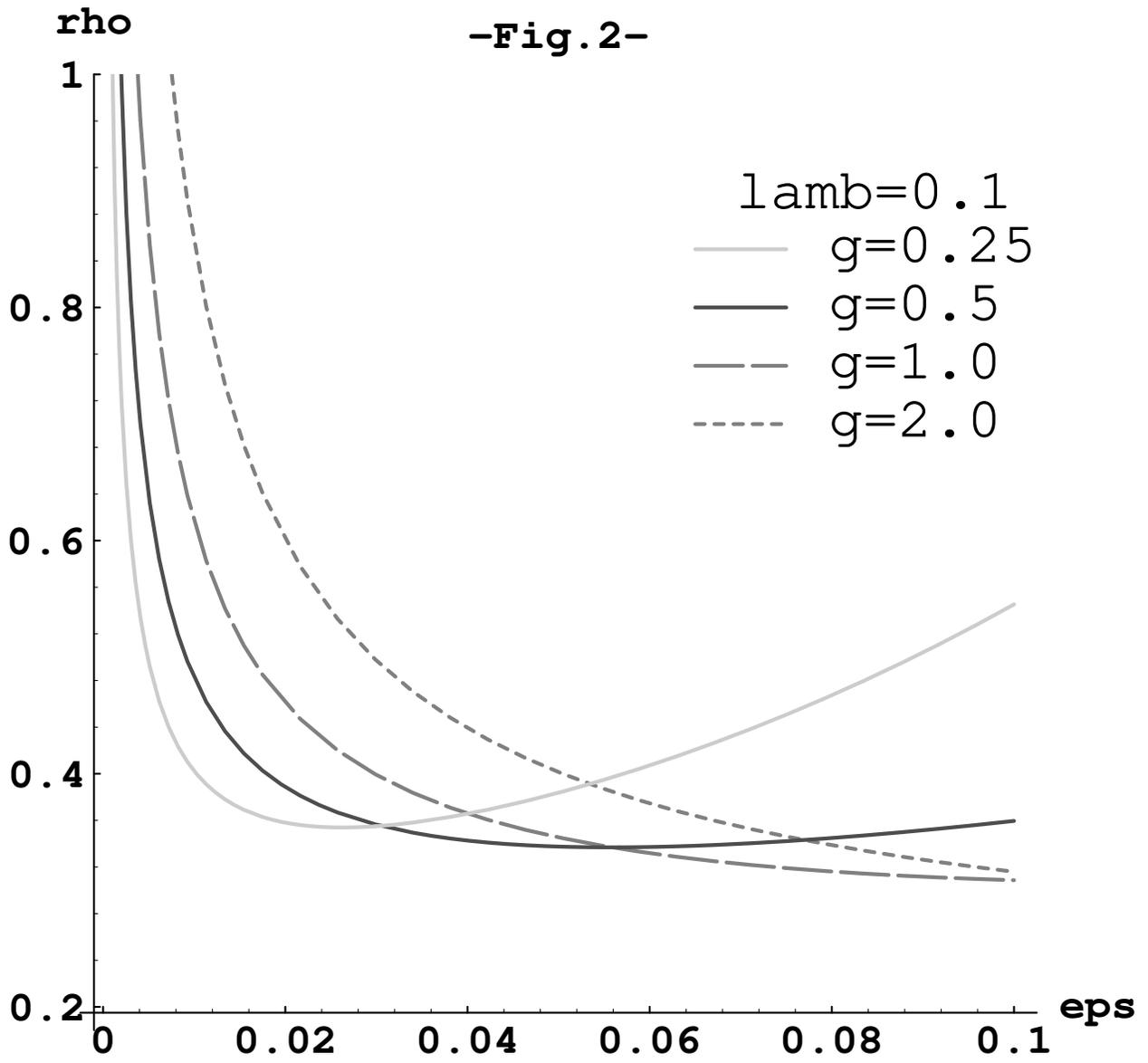
Fig. 2 : The density of states whose $g$ is varied from $0.25$ 
to $2.0$ at $\lambda=0.1$. 

\end{document}